\documentclass[prl,nofootinbib,twocolumn,superscriptaddress]{revtex4}
\usepackage{graphicx}

\def\mysection#1{{\bf #1.} }

\def\lsim{\mathrel{\rlap{\lower4pt\hbox{\hskip1pt$\sim$}}
    \raise1pt\hbox{$<$}}}         
\def\gsim{\mathrel{\rlap{\lower4pt\hbox{\hskip1pt$\sim$}}
    \raise1pt\hbox{$>$}}}         

\newcommand{\mP}{{\bar M}_{P}}

\begin{document}

\begin{titlepage}

\hfill$\vcenter{
\hbox{\bf SU-4252-843} }$

\hfill$\vcenter{
\hbox{\bf BNL-HET-07/3} }$

\hfill$\vcenter{
\hbox{\bf YITP-SB-07-02} }$

\begin{center}
{\Large \bf Warped Gravitons at the LHC and Beyond}
\vskip .1in
{\bf Kaustubh Agashe}$^*$,
{\bf Hooman Davoudiasl}$^\#$,\\
{\bf Gilad Perez}$^\dagger$
and {\bf Amarjit Soni}$^\#$\\
{\em $^*$ Department of Physics,
Syracuse University,
Syracuse, NY 13244}\\
 {\em $^\dagger$ C.N. Yang Institute for Theoretical Physics,
State University of New York, Stony Brook, NY 11794-3840}\\
{\em $^\#$ Brookhaven National Laboratory,
Upton, NY 11973}

\end{center}

\begin{center} {\bf Abstract}\\\end{center}

We study the production and decay of Kaluza-Klein (KK) gravitons at
the Large Hadron Collider (LHC), in the framework of a warped extra
dimension in which the Standard Model (SM) fields propagate.  Such a
scenario can provide solutions to both the Planck-weak hierarchy
problem and the flavor puzzle of the SM.
In this scenario, the
production via $q \bar{q}$ annihilation and decays to the conventional
photon and lepton channels are
highly suppressed. However, we
show that graviton production via gluon fusion
followed by decay to longitudinal $Z/W$ can be significant; vector boson
fusion is found to be a sub-dominant production mode.
In particular, the ``golden''  $ZZ$ decay mode offers a distinctive
4-lepton signal that could lead to the observation at the LHC with 300 
fb$^{-1}$ (SLHC with 3 ab$^{-1}$) of a
KK graviton with a mass up to $\sim 2$ ($\sim 3$) TeV
for the ratio of the AdS$_5$ curvature to the Planck scale
modestly above unity. We argue that (contrary to the lore)
such a size of the curvature scale
can still be within the regime of validity of the framework.
Upgrades beyond the SLHC luminosity 
are required to discover gravitons
heavier than $\sim 4$ TeV, as favored by the electroweak
and flavor precision tests in the simplest such models.

\vskip .2in

\end{titlepage}
\newpage
\renewcommand{\thepage}{\arabic{page}}
\setcounter{page}{1}

\section{Introduction}

Solutions to the Planck-weak hierarchy problem of
the Standard Model (SM) invoke new particles
at $\sim$ TeV scale.
Such new physics is likely to give a signal at the upcoming LHC, provided
that the new states have a non-negligible coupling to the
SM particles. In this paper, we consider the solution to
the hierarchy problem based on the Randall-Sundrum (RS1) framework with a
warped extra dimension \cite{rs1}. The most distinctive novel feature
of this scenario is the existence of spin-$2$ Kaluza-Klein (KK) gravitons
whose masses and couplings to the SM are set by the TeV scale. Hence, the KK
gravitons appear in experiments as widely separated resonances, in contrast to
the very light, closely spaced KK gravitons in
large extra dimensions \cite{add}, with couplings
suppressed by the 4-$d$ reduced Planck scale $\mP$.

A well-motivated extension of the original RS1 model addresses the
flavor structure of the SM through localization of fermions in the
warped bulk.  This picture offers a unified geometric explanation of
both the hierarchy and the flavor puzzles, without introducing a
flavor problem.  In this case,
graviton production and decay via light fermion channels are highly
suppressed and the decay into photons are negligible.  However,
finding a graviton spin-2 resonance provides the clearest evidence
for a warped extra dimension, on which the RS1 model and its
extensions are based.  The experimental verification of this
framework is then seemingly a challenge, since some of the most
promising original signals are no longer available.

Hence, we examine alternative LHC signals for RS1 KK gravitons,
assuming the SM fields are in the warped bulk and that the fermions
are localized to explain flavor. We show that production of KK
gravitons from gluon fusion and their decay into longitudinal gauge
bosons $W/Z$ $(W_L/Z_L)$ can be significant.  In particular, KK
graviton decay into pairs of $Z_L$'s can provide a striking 4-lepton
signal for a TeV-scale KK graviton at the LHC; multi-TeV gravitons
are shown to be accessible to a luminosity-upgraded LHC.  We also consider KK
graviton production via Vector Boson Fusion (VBF).
However, we find that this production channel is sub-dominant to that from
gluon fusion.  Hence, we do not analyze the VBF case in any detail.

\section{Warped Extra Dimension}

The framework is based on a slice of AdS$_5$. Owing to the warped geometry,
the relationship between the $5D$ mass scales
(taken to be of order $\mP$) and those in an
effective $4D$ description depends on the location in the extra dimension.
The $4D$ (or zero-mode) graviton is localized near the ``UV/Planck''
brane which has a Planckian fundamental scale, whereas
the Higgs sector is localized near the ``IR/TeV'' brane where it
is stable near a warped-down fundamental scale of order TeV. This
large hierarchy of scales can be generated via a modest-sized radius of
the $5^{\rm th}$ dimension: $\hbox{TeV} / \mP
\sim e^{ - k \pi R }$, where
$k$ is the curvature scale
and $R$ is the proper size of the extra dimension; $k R \approx 11$.
Furthermore, based on the AdS/CFT correspondence
\cite{Maldacena:1997re}, RS1
is conjectured to be dual to $4D$ composite Higgs models
\cite{Arkani-Hamed:2000ds, Contino:2003ve}.

In the original RS1 model, the entire SM (including the fermions
and gauge bosons) are assumed to be localized on the TeV brane.
The key feature of
this model is that KK gravitons have a mass $\sim$ TeV
and are localized near the TeV brane so that KK graviton coupling
to the {\em entire} SM is only $\sim$ TeV suppressed. Hence,
KK graviton production
via $q \bar{q}$ or $gg$ fusion at the LHC [or via
$e^+ e^-$ at International Linear Collider (ILC)] followed by
decays to dileptons or diphotons gives striking signals \cite{Davoudiasl:1999jd}.

However, in this model, the higher-dimensional operators in the $5D$
effective field theory (from cut-off physics) are suppressed only by
the warped-down scale $\sim$ TeV, giving too large contributions to
flavor changing neutral current (FCNC) processes 
and observables related to SM electroweak precision
tests (EWPT). Moreover, this set-up provides no understanding of the
flavor puzzle, {\it i.e.}, the hierarchies in the SM fermion Yukawa
couplings to Higgs.

An attractive solution
to this problem is to allow
the SM fields to propagate in the extra dimension \cite{bulkgauge, gn, 
Chang:1999nh, gp}.
In such a scenario, the
SM particles are identified with the zero-modes of the $5D$ fields
and the
profile of a SM fermion in the extra dimension
depends on its $5D$ mass parameter.
We can
then choose to localize 1st and 2nd generation fermions near the Planck brane
so that the
FCNC's from higher-dimensional operators are suppressed
by scales $\gg$ TeV which is the cut-off
at the location of these fermions~\cite{gp, hs}.
Similarly, contributions to EWPT from cut-off
physics are also suppressed.

As a bonus, we obtain a solution to the flavor puzzle
in the sense that hierarchies in the SM Yukawa couplings arise without
introducing hierarchies
in the fundamental $5D$ theory~\cite{gn, gp, hs}:
the 1st/2nd generation fermions
have small Yukawa couplings to Higgs which is localized near the
TeV brane.
Similarly,
the top quark can be localized near the TeV brane
to account for its large Yukawa.

On the flip side, in this scenario, couplings of KK gravitons to
light fermions are
highly suppressed
%
%
since, as mentioned above, KK gravitons are localized near TeV brane
whereas the light fermions are localized near the Planck brane.
In fact, we can show that these couplings
(made dimensionless by compensating the derivative involved
by $\sim$ TeV scale) are
very roughly of Yukawa strength
since KK gravitons have a
profile which is similar to that of the Higgs.
As a result,
$q \bar{q}$ annihilation at hadron collider
(or $e^+ e^-$ at ILC) to KK graviton is negligible.  
In contrast, SM gluons have a flat profile so that coupling to KK graviton
is suppressed only by a factor of the
%
%
size
of the extra dimension (in units of radius
of curvature), i.e., $k \pi R$,
relative to gluons being on the TeV brane. This factor
is basically $\sim \log \left( \mP /
\hbox{TeV} \right)$ due to solution to the hierarchy problem. Thus,
although suppressed compared to the original RS1 model,
the coupling of gluons to KK gravitons and hence
KK graviton production via $gg$ fusion is
still non-negligible (cf. the case of light fermions).

Furthermore,
decays of KK graviton are dominated by top quark and Higgs
due to their profile
being near TeV brane, resulting in couplings to KK gravitons
(which are also localized there) being
only
$\sim$ TeV-suppressed just like in the original RS1 model.
The problem
is that none of these
are easily detectable modes. Just as
with production of KK graviton, the branching ratio
(BR) to the usual golden modes, such as a pair
of photons, is volume suppressed, whereas
to light fermions is Yukawa-suppressed and hence negligible.
Thus, a priori, combination
of these 2 factors -- suppression
in production and in decays to the previously considered ``golden'' modes -- makes signal
for the KK graviton very difficult \cite{Davoudiasl:2000wi,
Davoudiasl:2001uj,
Davoudiasl:2003me}.

The crucial point of our paper
is that, by the equivalence theorem, $W^\pm_L$ and $Z_L$ are
effectively the {\em un}physical Higgs
(``would-be'' Goldstone bosons) and are therefore localized near
the TeV brane (just like the physical Higgs). 
So, the decay widths in the $W_L/Z_L$
channels are the same size as in those 
of the physical Higgs/top quark.\footnote{The longitudinal channels are dominant 
compared to those of the transverse $W/Z$ or gluon/photon by a volume
factor: in this sense, massive gauge bosons are different from the massless ones.}
Clearly, branching ratio to a pair of $Z/W$'s is sizable; in particular,
$Z_LZ_L$ is a golden channel. As a corollary,
{\em production} of KK graviton via longitudinal $W/Z$ fusion
can be important. Such
%
%
effects were not analyzed before.

Next, we comment on mass scale of KK gravitons.
In this scenario, there are new contributions to EWPT and FCNC's
calculable in the $5D$ effective field theory (EFT) from KK modes.
Due to various symmetries (approximate flavor or analog of GIM mechanism of
the SM \cite{gp, hs, aps}
and custodial isospin \cite{custodial}),
gauge KK masses as small as $\sim 3$ TeV
are consistent with oblique electroweak (EW) data \cite{custodial} (we comment on
non-oblique effects such as $Zb \bar{b}$ later) and FCNC's \cite{NMFV}.
As a result, KK gravitons have to be heavier than $\sim 4$ TeV
since
the ratio of lightest KK masses for graviton and gauge
bosons is $\sim 1.5$ in the simplest such models (see next section).

\section{Couplings of KK graviton}

A general
%
%
formula for
couplings of $m^{ \hbox{th} }$ and
$n^{ \hbox{th} }$ modes of the bulk field (denoted by $F$)
to the $q^{ \hbox{th} }$ level KK gravitons (denoted by $G$)
is \cite{Davoudiasl:2000wi}:
\begin{eqnarray}
{\cal L}_G & = & \sum_{ m , n, q } C^{ F F G }_{ m n q } \frac{1}{ \mP }
\eta^{ \mu \alpha } \eta^{ \nu \beta } h^{ ( q ) }_{ \alpha \beta }
( x ) T_{ \mu \nu }^{ ( m, n ) } ( x )
\end{eqnarray}
where $h^{ q }_{ \alpha \beta }
( x )$ corresponds to the KK graviton, $T_{ \mu \nu }^{ ( m, n ) } ( x )$
denotes the $4D$ energy-momentum tensor of the modes of the bulk field,
$\mP \approx 2.4 \times 10^{18}$ GeV
is the reduced $4D$ Planck scale
and $C^{ F F G }_{ m n q }$ is
the overlap integral of the wavefunctions of the $3$ modes.

We will consider only those couplings relevant for
production and decay.
Since $q \bar{q}$ annihilation to
KK graviton is Yukawa-suppressed, the production is dominated
by gluon fusion. The coupling of gluons
to KK gravitons is given by the above formula with \cite{Davoudiasl:2000wi}:
\begin{eqnarray}
C^{ A A G }_{ 0 0 n } & = &
e^{ k \pi R }
\frac{ 2 \left[ 1 - J_0 \left( x_n^G \right) \right] }{ k \pi R
\left( x_n^G \right)^2 | J_2 \left( x_n^G \right) | }
\end{eqnarray}
where $J_{ 0, 2 }$ denote Bessel functions
and $x^G_n = 3.83, 7.02, 10.17, 13.32$ gives masses of
the first 4 KK gravitons: $m^G_n = k e^{ - k \pi R } x^G_n$. Gauge KK masses
are given by
$m^A_n = k e^{ - k \pi R } \times ( 2.45, 5.57, 8.7, 11.84 )$.
For simplicity, we neglect brane-localized kinetic terms
for both graviton and gauge fields.
Thus, we have
\begin{eqnarray}
m^G_1 & \approx 1.5 m^A_1,
\label{KKgKKG}
\end{eqnarray}
for the lightest KK masses for graviton and gauge fields.

As mentioned above, the decays of
KK graviton are dominated by top
quark and Higgs (including longitudinal $W/Z$
using equivalence theorem). Let us consider the
top and bottom sector in detail to determine
the couplings to KK graviton. Due to heaviness of top quark
combined with constraint from shift in $Z b \bar{b}$,
one
possibility is to
localize $t_R$ very close to TeV brane with $(t,b)_L$ having
a profile close to flat \cite{custodial}.
Even with this choice of the profiles, the gauge KK mass scale is constrained
by $Z b \bar{b}$ to
be $\stackrel{>}{\sim} 5$ TeV, i.e., a bit higher than that allowed
by oblique EW data.  However,
a
%
%
custodial symmetry to suppress $Z b \bar{b}$ \cite{Zbb}
can relax this constraint on the
gauge KK mass scale and moreover allows
the other extreme case: $(t,b)_L$ very close to the TeV brane
and $t_R$ close to flat
and also the intermediate possibility with both
$t_R$ and $(t,b)_L$ being near, but not too close to
TeV brane.
The bottom-line is that,
with this custodial symmetry and for certain choices of profiles
for $t_R$ and $(t,b)_L$ in the extra dimension, gauge KK masses as low as
$\sim 3$ TeV can be consistent with $Z b \bar{b}$
as well \cite{Carena:2006bn}.
For simplicity, we will consider the extreme
case with
%
%
$t_R$
%
%
localized very close to the
TeV brane, with $(t,b)_L$ having close to a flat profile.
It is straightforward to extend our analysis
to the other cases.
Moreover, we will assume that
this helicity of the top quark and similarly the Higgs
are {\em exactly} localized on the TeV brane.
In reality, these particles have a profile {\em peaked}
near the TeV brane, but this will result in
at most an $O(1)$ difference.

With this approximation, the couplings relevant for
decay
are:
\begin{eqnarray}
{\cal L}_G & \ni & \frac{e^{ k \pi R } }{ \mP }
\eta^{ \mu \alpha } \eta^{ \nu \beta } h^{ ( q ) }_{ \alpha \beta }
( x ) T_{ \mu \nu }^{ t_R, H } ( x )
\end{eqnarray}
giving the partial decay widths \cite{Han:1998sg}:
\begin{eqnarray}
\Gamma \left( G \rightarrow
t_R \bar{ t_R } \right) & \approx & N_c \frac{ (c\, x^G_n)^2 \, m^G_n }
{ 320 \pi } \\
\Gamma \left( G \rightarrow h h \right) & \approx &
 \frac{ (c\, x^G_n)^2 \, m^G_n }
{ 960  \pi } \\
\Gamma \left( G \rightarrow W^+_L W^-_L \right) & \approx &
 \frac{ (c\, x^G_n)^2 \, m^G_n }
{ 480  \pi } \\
\Gamma \left( G \rightarrow Z_L Z_L  \right) & \approx &
 \frac{ (c\, x^G_n)^2 \, m^G_n }
{ 960  \pi }
\end{eqnarray}
where $N_c = 3$ is
number of QCD colors, $c\equiv k/\mP$,
and we have neglected
masses of final state particles
in phase space factors.  These are the only important decay
channels for the $n=1$ graviton KK mode which is the focus of
our analysis in this work.
For the case where
$(t,b)_L$ is localized very close to the
TeV brane (with $t_R$ being close to flat),
we multiply 1st formula by a factor of $2$ to include decays
to $b_L$.
In this case,
production of KK graviton from $b \bar{b}$ annihilation
can also be important.
The last 2 formulae correspond to decays to longitudinal
polarizations: we have used
equivalence theorem
(which is valid
up to $M_{ W, Z }^2 / E^2$ effects, where $E \sim m^G_1$) to relate
these decays to physical Higgs.
As mentioned above, we can
neglect decays to transverse $W/Z$ (and similarly to gluon, photon)
due to volume
[$\sim \log \left( \mP
/ \hbox{TeV} \right)$] suppression (in amplitude)
relative to longitudinal polarization. Similarly, decays
to light fermions are negligible (due to the Yukawa-suppressed
coupling to KK graviton).
We can also show that the decays of KK graviton
to other KK modes are suppressed.
Finally, for the intermediate possibility mentioned above
(with both
$t_R$ and $(t,b)_L$ being near, but not too close to
TeV brane), the partial width of KK graviton
to top/bottom quarks (and hence the total width) will be smaller and
hence the
BR to $ZZ$ will be larger.

\section{KK graviton production}

The relevant matrix elements for the process
$gg \rightarrow VV$, with $V=W,Z$, via KK graviton are
\cite{Park:2001vk}:
\begin{eqnarray}
{\cal M}^G_{ \lambda_1 \lambda_2 \lambda_3 \lambda_4 }
\left( g^a g^b \rightarrow VV \right) &=&
- C^{ A A G }_{ 0 0 n } e^{ - k \pi R }
\left( \frac{ x^G_n c }{ m^G_n } \right)^2
%
%
\nonumber \\
 & \times & \sum_n \frac{ \delta_{ a b }\,
[{\cal A}_{ \lambda_1 \lambda_2 \lambda_3
\lambda_4 }]}{ \hat{s} - m_n^2 + i
\Gamma_G m_n }
\nonumber \\
\label{MG1}
\end{eqnarray}
where $\lambda_i$ refer to initial and final state polarizations,
$a, b$ are color factors,
\begin{eqnarray}
\Gamma_G & = & \frac{ 13 (c\, x^G_n)^2 \, m^G_n }{ 960 \pi}
\label{width}
\end{eqnarray}
is the total decay
width of KK graviton in our treatment, and we have used
$\mP\, e^{ - k \pi R } = m^G_n/(x^G_n c)$.  As mentioned before, $x^G_1 = 3.83$
for the first graviton resonance.  We have
\begin{eqnarray}
{\cal A}_{ + + 0 0 } & = & {\cal A}_{ - - 0 0 } = 0
\nonumber \\
 & &
\\
{\cal A}_{ + - 0 0 } & = & {\cal A}_{ - + 0 0 } \nonumber \\
& = &
\frac{
\left( 1 - 1 / \beta_V^2 \right)
\left( \beta_V^2 - 2 \right)
\Big[ \left( \hat{t} - \hat{u} \right)^2
- \beta_V^2 \hat{s}^2 \Big] \hat{s} }{ 8 M_V^2 } \nonumber \\
\label{ggfusion}
\end{eqnarray}
where $\beta_V^2 = 1 - 4 M_V^2 / \hat{s}$ 
and the hatted variables are in the parton center of mass frame.  To repeat,
the other amplitudes with transverse polarizations for
$V$'s (i.e.,
$\lambda_{3,4} = +, -$)
can be neglected since these are suppressed relative
to the above by $\sim \log \left( M_{ Pl } / \hbox{TeV} \right)$.
Note that the above formula
includes both the virtual exchange of KK graviton and resonant
production.  One can show that 
${\cal A}_{ + - 0 0 } \to -\sin^2\hat{\theta} \, \hat{s}^2/2$ 
as $\beta_V \to 1$.

The parton-level signal ($V=Z$) cross-section, averaged over initial
state spins and colors, is given by:
\begin{eqnarray}
\frac{ d \hat{ \sigma } \left( g g \rightarrow ZZ \right) }{ d \cos
\hat{ \theta } }
& \approx &
\frac{| {\cal M}_{ +- 00 } |^2 }{ 1024 \pi \hat{s} }
\label{dsig/dc}
\end{eqnarray}
where a factor of $1/2$ has been included for identical bosons in the final state,
initial helicity averaging has been accounted for by a factor of $1/4$
and a factor of $1/8$ accounts for color averaging.
Note that ${\cal M}_{ +- 00 }$ is the only independent non-zero matrix element for the above process.
The total parton level cross section $\hat{\sigma}$ is related to the proton-level
total signal cross-section as usual:
\begin{eqnarray}
\sigma ( pp \rightarrow ZZ )
\!\!& = &
\!\!\!\int \!\!dx_1 dx_2 f_g \left( x_1, Q^2 \right)\!f_g \left( x_2, Q^2 \right)
\!\hat{ \sigma } \left(
x_1 x_2 s \right), \nonumber \\
\label{sigtot}
\end{eqnarray}
where $f_g$ are the gluon PDF's
and $Q^2 \sim ({m^G_n})^2$ is the typical momentum transfer
in the partonic process for resonant production
of a KK graviton.

Finally, we discuss a new production mechanism for KK graviton
which has not been considered before, namely, VBF via $WW$ or $ZZ$.
The probability for
emission of (an almost) collinear longitudinal $W/Z$ by
a quark (or anti-quark) is suppressed by
electroweak factor of
$\sim \alpha_{ EW } / \left( 4 \pi \right)$ \cite{Han:2005mu}. However,
the coupling of longitudinal $W/Z$ to KK graviton
%
%
is
$\log \left(\mP / \hbox{TeV} \right)$-enhanced
compared to that to gluon (or to transverse $W/Z$).
Moreover, VBF
can proceed via valence quarks, i.e., $u u$ or $u d$, scattering
in addition to $u \bar{u}$ and $d \bar{d}$ annihilation
(which are suppressed
by the smaller sea quark content). So, we find that the ratio
of KK graviton production via longitudinal $W/Z$ fusion and
gluon fusion is $\sim \Big[ \alpha_{ EW } / \left( 4 \pi \right) \Big]^2 \times
\Big[ \log \left( \mP / \hbox{TeV} \right) \Big]^2
\times$ ratio of ($u$ PDF)$^2$ vs.($g$ PDF)$^2$.
Since ($u$ PDF)$^2$ is roughly an order of magnitude larger
than ($g$ PDF)$^2$ at the relevant $x$'s,
%
%
we
estimate that the cross-section for $gg$ fusion
is about an order of magnitude larger than the
$WW/ZZ$ fusion -- our detailed, partonic level, calculation confirms
this expectation. Further details are discussed in the appendix.
%
%

\section{SM background}

In the next section,
we focus mainly on the leptonic decay mode of the two $Z$s (4$\ell$),
based on considerations of background as we now discuss.
We begin with the irreducible background to
the $ZZ$ final state, i.e., SM contribution
to $pp \rightarrow ZZ + X$.
%
%
%
%
%
It is dominated by $q \bar{q}$ annihilation:
gluon fusion is very small in the SM since it
proceeds via loop. Hence, the interference
of KK graviton signal (dominated by
$gg$ and $WW/ZZ$ fusion) with the SM background is negligible.
The parton-level cross-section, averaged
over quark colors and spins is given by \cite{EHLQ}
\begin{eqnarray}
\frac{ d \hat{ \sigma } \left( q_i \bar{q}_i \rightarrow ZZ \right) }
{ d \hat{t} } & = & \frac{ \pi \alpha^2 \left( L_i^4 + R_i^4 \right) }
{ 96 \sin^4 \theta_W \cos^4 \theta_W \hat{s}^2 }  \nonumber \\
& \times&
\Big[ \frac{ \hat{t} }{ \hat{u} } + \frac{ \hat{u} }{ \hat{t} } +
\frac{ 4 M_Z^2 \hat{s} }{ \hat{t} \hat{u}}
- M_Z^4 \left( \frac{1}{ \hat{t}^2 } + \frac{1}{ \hat{u}^2 } \right) \Big],
\nonumber \\
\end{eqnarray}
where $L_u = 1 - 4/3 \sin^2 \theta_W$, $R_u = - 4/3 \sin^2 \theta_W$,
$L_d= -1 + 2/3 \sin^2 \theta_W$ and $R_d = + 2/3 \sin^2 \theta_W$
This cross-section exhibits
forward/backward peaking due to $t/u$ channel exchange, whereas
KK graviton signal does not have this feature
[see the approximate
$\hat \theta$ dependencies given below Eqs. (\ref{ggfusion} and \ref{VBF})]. Hence,
a
cut on pseudo-rapidity $\eta$ is useful to
reduce this background keeping the
signal (almost) unchanged. Thus as we shall see below our signal
is typically significantly larger than the SM
$ZZ$ background.

The smallness of
the irreducible background to $ZZ$ final state leads us to consider
the reducible background which depends on the decay mode of the $Z$ pair.
For the dominant
purely hadronic decay mode, there is a huge QCD background
($4$ jets) so that this decay mode is not useful.
Next , we consider the semi-leptonic decay mode.
The problem is that for such energetic
$Z$'s, the opening angle between
$2$ jets from $Z$ decay $\sim M_Z / 1 \;
\hbox{TeV} \sim 0.1$, whereas the typical cone size
for jet reconstruction is $\sim 0.4$ (see for example \cite{conejetalgorithm}).
Hence, it is likely that we cannot resolve the
$2$ jets from $Z$ decay so that they will appear as a single jet
(``$Z$-jet'').
%
%
Therefore, we need to consider the
%
%
background
from $Z + 1$ jet which we calculate
is 
roughly 
%
%
an order of magnitude
larger than our signal (over the same mass window)
-- note that, based on the above discussion,
this statement is true irrespective of the value of $c$.
However, we note that
more sophisticated
means of reducing the $Z + 1$ jet background,
for example via a better set of cuts
or by looking for a substructure
inside the $Z$-jet from KK graviton decay (this will give a hint that the jet is neither
a light jet nor a b-jet),
might make this channel useful.

Also, VBF has the feature of
$2$ additional highly energetic
forward-jets which can be tagged \cite{Han:2005mu}.
However in this case and with semi-leptonic decay
of $Z$ pair, we will have to consider
background from $Z + 3$ jets,
with its associated
QCD uncertainties.
Moreover, VBF is sub-dominant to $gg$ fusion
and hence VBF might not have enough statistics
(for the interesting range of KK masses) which are
required for an analysis involving forward-jet tagging.
In view of these difficulties
with the semi-leptonic decay mode, here we will
follow a conservative approach and not consider this decay mode,
but we note that it is worthy of a future study.
So, for now, we will focus on the
purely (charged) leptonic decay mode for $ZZ$ for which the dominant
background is the irreducible one and hence is smaller than our signal.
%
%
The channel with one $Z$ decaying
to neutrinos, whereas the other $Z$ decays to charged leptons
is also interesting. The
BR for this channel is larger than for the
$4 l$ mode, but
the invariant mass of the
$Z$ pair cannot be reconstructed in this case
so that we cannot apply the
mass window cut (see below) to enhance the ratio of signal over
background. However, the distribution of
kinematic variables such as
missing $p_T$ will still be different for signal as compared to 
the SM 
background which will help in discriminating between
the two -- we will defer this analysis for
a future study.

\section{Signals at the LHC}

\begin{figure}
\includegraphics[width=0.48\textwidth]{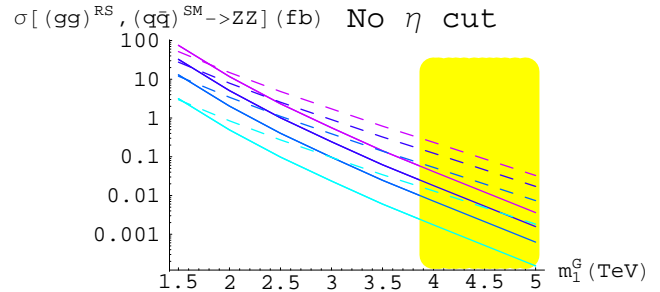}
\caption{The cross-sections (integrated over one width) for
$gg \rightarrow ZZ$ via
KK graviton (solid lines) and the corresponding SM background (dashed lines).
We show the cross-sections for $c \equiv k / \mP =
0.5, 1, 1.5, 2$ (from bottom to top). See the text for an explanation
of the upper limit on $c$.
The yellow region shows where we expect the KK graviton mass to be
in the simplest models according to relation in Eq~\ref{KKgKKG}
and the limit on gauge KK mass from precision tests.}
\label{nocuts+b}
\end{figure}

\begin{figure}
\includegraphics[width=0.48\textwidth]{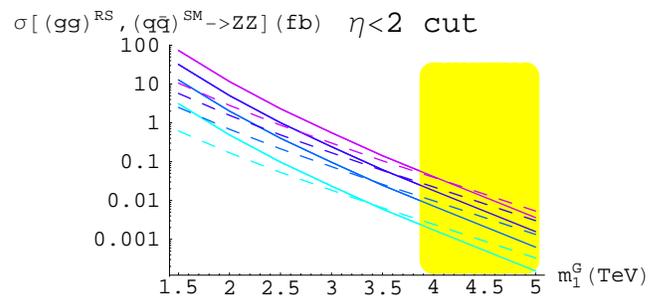}
\caption{Same as fig.~\ref{nocuts+b}, but with $\eta<2$.}
\label{cuts+b}
\end{figure}

Our results for the  KK graviton signal (S) and irreducible SM
background (B) cross sections,
both within the $ZZ$ invariant mass
window \footnote{The
ratio of signal to background is maximized when KK graviton
is on-shell and, therefore, we focus on this region.} $m^G_1 \pm
\Gamma_G$,
are presented
in Figs.~(\ref{nocuts+b}) and (\ref{cuts+b}), 
without a cut on eta and with such a
cut, respectively.
%
%
The yellow region shows where we expect the KK graviton mass to be
in the simplest models according to relation in Eq.~\ref{KKgKKG}
and the limit on gauge KK mass from precision tests.
As expected from the above discussion, these results show that implementing a
cut with $\eta < 2$ on the final state $Z$'s enhances $S/B$.
With this cut on $\eta$
we find that the signal
is larger than the background by a factor of a few
(or even an order of magnitude) over a wide range of KK
graviton masses.
We consider the
range $0.5 \leq c\leq 2$; a discussion of the upper limit
on $c$ is given later.
We have used the
CTEQ6L1 PDF's, evaluated at $Q^2 = m^{ G \; 2}_1$ in our (partonic level)
calculations (both for the signal and the background).
Note that, based on
Eqs. (\ref{MG1}), (\ref{width})
and (\ref{MG2}),
the dependence on
$c$ parameter (roughly) cancels in
parton-level signal cross-section
(or equivalently the proton-level {\em differential}
cross-section)
{\em near the peak
of the resonance}. Hence, the ratio of signal to background
in this mass window is (almost) independent of $c$.


Based on the preceding discussion, we are led to consider the
purely leptonic ($e^+e^-, \, \mu^+\mu^-$) decay of the $Z$ pair.
However, this decay mode
has a small
BR of $\approx 0.45 \%$.
%
Hence, the main issue
is whether the
number of $4\ell$ events from signal and also
the $S / \sqrt{B}$ is large enough,
especially given the small BR of this mode.
The {\em total}
%
%
signal cross-section and hence the number of $4\ell$ events
scales as $c^2$. The reason is
that (as explained above) the {\em differential} cross-section
at the peak of the resonance is (roughly) independent of
$c$, but the size of the mass window $\sim \Gamma_G \propto c^2$.
Similarly, $S/ \sqrt{B}$ scales as $c$.

{\bf Range of $c$}. Therefore, it becomes crucial to study the
allowed range of $ k / \mP$. Recall that there is an upper limit on
$c$ such that the assumption of neglecting higher curvature terms is
valid \cite{rs1, Davoudiasl:2000wi}. The common lore is that $c \sim
1$ is outside the domain of validity of the model. However, we now
show that $c \sim 1$ is still {\em within} the range of validity of
the model. The point is that the higher curvature terms in the $5D$
action are suppressed by powers of $R_5 / \Lambda^2$, rather than $R_5/
M_5^2$, with $R_5 = 20 k^2$ the size of the $5D$ curvature 
\cite{Davoudiasl:2000wi} and $M_5$
the $5D$ Planck scale. Here $\Lambda$ is the
energy scale at which the $5D$ gravity theory becomes strongly
coupled and its Naive Dimensional Analysis (NDA) estimate is given
by $\Lambda^3 / \left( 24 \pi^3 \right) \sim M_5^3$
\cite{Chacko:1999hg} . We can show that loop effects and local
higher-dimensional operators in the $5D$ theory are also suppressed
by a similar factor. Using the relation $ \mP^2 \approx M_5^3 / k$,
we require $k / \mP < \sqrt{ 3 \pi^3/(5\sqrt{5})}$ so that we can trust our
above calculation of the tree-level effects of KK gravitons.
Although there are $O(1)$ uncertainties in these estimates, we thus
expect that
%
%
for $k / \mP \sim 1$, higher-order
corrections to our results can be neglected.
In fact,
even
$c$ modestly
%
%
larger than $1$
can still be within the regime of validity
of the model since
the edge of validity of the model is $k / \mP \approx 3$.
Hence, we will consider values of $c$ as large as $2$
in our results.\footnote{Note that for
values of $c$ larger than $\sim 2$, the KK graviton width becomes larger
than $\sim 20 \%$ of its mass, making some of the approximations
used in our calculations less reliable and also
introducing additional detection issues.}

{\bf Other decay modes of KK graviton}.
Before presenting our results based on the $4\ell$ events,
we would briefly like to mention other decay modes of the KK graviton,
beginning with the dominant decay mode to top quarks
(BR $\approx 70 \%$).
The
purely leptonic decay mode for the top pair
(i.e.,  $W$'s from both tops decaying leptonically) has very small BR
($\approx 5\%$) and hence is too inefficient. The
semi-leptonic decay mode has large BR ($\approx 30 \%$), but
it was shown in reference \cite{kkgluon}
that for $p_T$ of top quark $\stackrel{>}{\sim} 1$ TeV
(as would be the case for KK graviton masses of interest),
the $C4$ jet algorithm \cite{conejetalgorithm} is unable to resolve the
$3$ jets from hadronic top decay ($b$-jet and $2$ jets from $W$ decay),
just like for the case of hadronic decay of $Z$ mentioned above.
Hence the conventional hadronic top
reconstruction methods for
$t \bar{t}$ invariant
masses $\stackrel{<}{\sim} 600$ GeV \cite{oldttbar}
are inefficient for such energetic tops.
The new methods proposed in reference \cite{kkgluon},
based on this
``top-jet'',
results in a total
efficiency of $\sim 1 \%$ (including BR, $b$-tagging
efficiency
and kinematic effects)
for the case of a $3$ TeV KK {\em gluon} decaying into top
pairs
\cite{kkgluon}. We expect a similar small efficiency
for KK graviton masses $\stackrel{>}{\sim} 2$ TeV.
The case of
decays of KK graviton to $WW$ followed by
leptonic decays of both $W$ is also problematic,
since the neutrinos' $p_T$ will tend to be almost back to back, due to the high boost of the $W$'s.
Thus in many cases the missing energy information will be lost and the $W$ mass cannot be reconstructed efficiently.
Of course the hadronic decay of $W$ faces the same problem as above
for top/$Z$ hadronic decay.
Thus, we conclude that the other decay modes of KK graviton
might be
more challenging and less clean that the $4\ell$ mode we are considering,
but these other decay modes certainly
deserve a separate and more detailed study
(especially the decays to top quarks since the top
decays, in turn, carry useful spin information).

\begin{figure}
\includegraphics[width=0.48\textwidth]{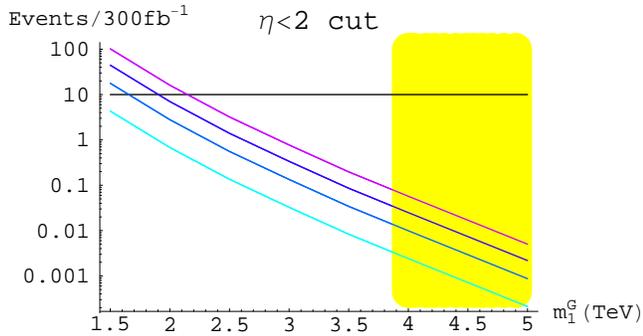}
\caption{The total number of expected events for the purely
leptonic decay mode for $Z$ pairs from
KK graviton decay using 300~fb$^{-1}$ with $\eta<2$. 
See also Fig.~(\ref{nocuts+b})}
\label{totalN}
\end{figure}

\begin{figure}
\includegraphics[width=0.48\textwidth]{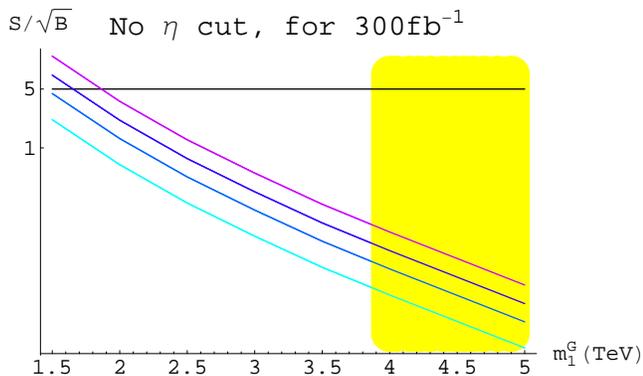}
\caption{Significance for the purely leptonic decay mode
for $Z$ pairs from
KK graviton
using 300~fb$^{-1}$. 
See also Fig.~(\ref{nocuts+b})}
\label{StoRBnoet}
\end{figure}

\begin{figure}
\includegraphics[width=0.48\textwidth]{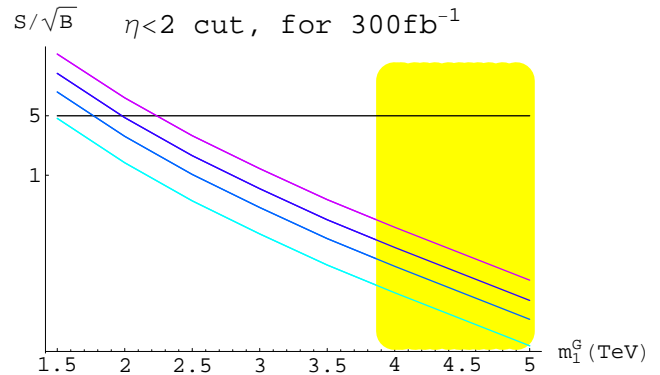}
\caption{Same as FIG.~\ref{StoRBnoet}, but
with $\eta<2$.}
\label{StoRB}
\end{figure}

{\bf Results}.
%
%
In Fig.~\ref{totalN},
%
%
we show the number of
$4\ell$ events for the LHC with $300$ fb$^{-1}$ luminosity
%
%
and in Figs.~\ref{StoRBnoet}
and \ref{StoRB}
we show the statistical
significance of the signal ($S / \sqrt{B}$), with and without the
$\eta$ cut, respectively -- we again see the
importance of the $\eta$ cut in improving the
significance of the signal.
We define the reach to be largest KK mass
for which  number
of $4\ell$ events $\geq 10$, provided also that $S / \sqrt{B} \geq 5$.
%
%
(We assume for simplicity 100\%
efficiency since our signal is known to be one of the cleanest at the LHC.)
In Table \ref{t1}, we show
this reach of the LHC
from which
we see that for $c \stackrel{<}{\sim}2$, the LHC can probe
KK graviton masses up to $\sim 2$ TeV.
Recall that the constraints from FCNC and EWPT
on gauge KK masses in the simplest {\em existing} models
in the literature require KK graviton masses
$\stackrel{>}{\sim} 4$ TeV.
%
%
%
%

We also note that higher luminosities of
$3 \; \hbox{ab}^{-1}$ are being
discussed in the community for the SLHC
(see for example references \cite{Gianotti:2002xx, Bruning:2002yh}).
The number of $4\ell$ events and $S/ \sqrt{B}$ for the SLHC can
be easily obtained by multiplying
the corresponding numbers for
the LHC by $10$ and $\sqrt{10}$, respectively.
From Table \ref{t2}, we see that the SLHC can extend the reach
for KK graviton
to $\sim 3$ TeV.
%
%
Similarly,
upgrades of the center of mass energy to
$28$ TeV (see for example reference \cite{Bruning:2002yh}) can extend the
reach in KK masses.
Note that the $4$-lepton signal is the cleanest
(in terms of background) of the possible
KK graviton decay modes.
This feature makes it a very promising discovery mode for KK graviton
even at higher luminosity/energy, cf.
other modes (including the dominant decay
mode to top quarks) which involve hadrons.

An alternate possibility
is that new model-building avenues or mechanisms
to suppress EWPT and FCNC
allow lower gauge (and hence graviton) KK
masses, just as
the custodial symmetries to suppress contributions
to $T$ parameter and $Z b \bar{b}$ coupling
relaxed the constraints on the KK masses
before.\footnote{For example, references
\cite{Hirn:2006nt}
discuss the possibility of suppressing
the $S$ parameter while keeping the solution
to the flavor puzzle intact.}
%
%

%
\begin{table}
[t]
\begin{tabular}{lcccc}
\hline \hline
$c\equiv k/\mP\quad\quad$
 & \hspace*{.2cm}$~~~~~~~0.5\quad\quad$
 &   $~~~~~~1.0\quad\quad$
 &   $~~~~~~1.5\quad\quad$
 &   $2.0$\\
\hline
$m^G_1$ (TeV)
 & $<1.5$
 &  $1.6$
 &  $1.9$
 &  $2.2$ \\
 \hline
$S/\sqrt{B}$
 &  $-$
 &  $7.0$
 &  $6.1$
 &  $6.1$ \\
\hline\hline
\end{tabular}
\caption{The mass of the first KK graviton for which the number of signal
events is 10 at the LHC, for various choices of $c$.
See the text for an explanation
of the upper limit on $c$. The significance
$S/\sqrt{B}$ of each result is also given.  These numbers correspond to
300~fb$^{-1}$ of integrated luminosity.
}
\label{t1}
\end{table}

\begin{table}
[b]
\begin{tabular}{lcccc}
\hline \hline
$c\equiv k/\mP\quad\quad\quad$
 &   $\,~~~~~0.5\quad\quad$
 &   $~~~~~1.0\quad\quad$
 &   $~~~~~1.5\quad\quad$
 &   $~~2.0$\\
\hline
$m^G_1$ (TeV)
 & $1.9$
 &  $2.3$
 &  $2.6$
 &  $2.9$ \\
 \hline
$S/\sqrt{B}$
 &  $6.1$
 &  $4.3$
 &  $4.3$
 &  $4.3$ \\
\hline\hline
\end{tabular}
\caption{Same as TABLE I, except for the SLHC
with 3~ab$^{-1}$ of integrated luminosity.
}
\label{t2}
\end{table}

Finally, it
is interesting that, although we
might not have enough statistics for
a few TeV KK graviton masses, the $Z/W$ pairs from KK graviton
can be discriminated from SM background as follows. First of all,
the (reconstructed) $Z/W$ pairs from KK graviton have a characteristic
spin-2 angular distribution as opposed to the SM background.
Also, the SM $ZZ$'s are mostly transverse, whereas the ones from KK graviton
are mostly longitudinal.
Hence, the angular distribution of decay products of $Z$
in the $Z$ {\em rest} frame (or their energy distribution
in the lab frame) can also distinguish KK graviton
signal from SM background.

\section{Conclusions}

In this work, we have studied the discovery potential,
at the LHC and its future upgrades, for the
first RS graviton KK mode, assuming bulk SM.  Such a
discovery will provide strong evidence in favor of the RS model as
the resolution of both the Planck-weak
and the flavor hierarchy puzzles.  We considered gluon-fusion and
VBF production modes and  found that the VBF mode is
sub-dominant.  We focused on a remarkably clean 4-lepton signal,
originating from the decay of the graviton to 2 longitudinal $Z$'s.
With this signal, the reach of the LHC for
the first graviton KK mode extends to around 2~TeV, for
an integrated luminosity of 300~fb$^{-1}$
and for the ratio of the AdS$_5$ curvature to $\mP$
modestly above unity, which as we argued (and contrary to the lore)
can still be within the regime of validity
for our computations.
On the other hand, within the (simplest) current theory understanding, the
electroweak and flavor precision tests disfavor
KK graviton masses below $\sim 4$ TeV.  However,
the discovery reach can be extended
at the upgraded SLHC luminosity of order 3~ab$^{-1}$ and
approach 3~TeV.
Finally, we discussed briefly how the semi-leptonic decay mode
of the $Z$ pairs from KK graviton can be useful
with a more
refined analysis designed to reduce the background.

{\it Note added:} While this work was being finalized,
Ref.~\cite{Fitzpatrick:2007qr} appeared containing a similar discussion,
in the context of bulk SM, of
the couplings of KK gravitons
to longitudinal $W/Z$, based
on the equivalence principle, but focusing on the search
for the KK graviton at the LHC using its decays to top quarks.

\mysection{Acknowledgments}
We thank Tao Han, Tadas Krupovnickas, Seong-Chan Park,
Maxim Perelstein, David
Rainwater, Tom Rizzo, and Joseph Virzi for useful discussions and
the Aspen Center for Physics for hospitality during
part of this project.
KA is supported in part by the U. S. DOE under
Contract no. DE-FG-02-85ER 40231.
HD and AS are supported in part by the DOE grant
DE-AC02-98CH10886 (BNL).

\section{Appendix}

The relevant matrix elements for the process $VV \rightarrow
ZZ$ via KK graviton exchange are given by:
\begin{eqnarray}
{\cal M}^G_{ \lambda_1 \lambda_2 \lambda_3 \lambda_3 }
\left( VV \rightarrow ZZ \right)  &=&
\left( \frac{ x^G_n c }{ m^G_n } \right)^2 \nonumber
%
%
\\
& \times &
\sum_n \frac{ {\cal A}_{ \lambda_1 \lambda_2 \lambda_3 \lambda_3 } }
{ \hat{s} - m_n^2 + i \Gamma_G m_n },
\label{MG2}
\end{eqnarray}
where $V = W, Z$ and
\begin{eqnarray}
{\cal A}_{ 0000 } & = &
\left( \beta_V^2 - 1 \right)
\left( \beta_V^2 - 2 \right) \left( \beta_Z^2 - 1 \right)
\left( \beta_Z^2 -2  \right) \nonumber \\
&\times&\frac{ \Big[ 3 \left( \hat{t} - \hat{u} \right)^2 - \hat{s}^2
\beta_V^2 \beta_Z^2
\Big] \hat{s}^2 }
{ 96 \beta_Z^2 \beta_V^2 M_Z^2 M_V^2 },
\label{VBF}
\end{eqnarray}
From the above discussion, it is clear that the other amplitudes
with transverse polarizations for initial or final state
bosons can be neglected due to the smaller
couplings to the KK graviton. We can show that in the limit
$\beta_{ W, Z } \to 1$,  $A_{ 0 0 0 0 } \rightarrow
\hat{s}^2 / 2 \left(  2/3 - \sin^2 \theta \right)$.

The parton-level cross-section is given by
\begin{eqnarray}
\frac{ d \hat{ \sigma } \left( V_L V_L
\rightarrow ZZ \right) }{ d \cos \hat{ \theta } }
& \approx & \frac{ | {\cal M}_{ 0 0 0 0 } |^2 }{ 64 \pi \hat{s} }
\end{eqnarray}
where the subscript $L$ on $V$ denotes longitudinal polarization.

The probability distribution
for a quark of energy $E$ to emit a longitudinally
polarized gauge boson
of energy $xE$ and transverse momentum
$p_T$ (relative to quark momentum) is approximated by
\cite{Han:2005mu}:
\begin{eqnarray}
\frac{ d P^L_{ V / f } \left( x, \; p_T^2\right) }{ d p_T^2 }
& = & \frac{ g_V^2 + g_A^2 }{ 4 \pi^2 } \frac{ 1 - x }{x}
\frac{ ( 1 - x ) M_V^2 }{ \Big[ p_T^2 + ( 1 - x )
M_V^2 \Big]^2 }
\nonumber \\
\label{Lemission}
\end{eqnarray}

The proton-level cross-section can then be written as
\begin{eqnarray}
\sigma \left( pp \rightarrow ZZ
\right) & \ni &
 \int d x_1 d x_2 d x_1^W d x_2^W d p_{ T \; 1 }^2
d p_{ T \; 2 }^2 \nonumber \\
& \times &
\frac{ d P^L_{ W/u }
\left( x_1^W, \; p_{ T \; 1 }^2 \right) }{ d p_{ T \; 1 } ^2 }
\frac{ d P^L_{ W/d }
\left( x_2^W, \; p^2_{ T \; 2 } \right) }{ d p_{ T \; 2 } ^2 }
\nonumber \\
& \times& f_u (x_1, Q^2 ) f_d (x_2, Q^2 ) \hat{ \sigma } \left( \hat{s} \right)
 \nonumber \\
 &+& (u \leftrightarrow d)\nonumber\\
& \approx &
\int d x_1 d x_2 d x_1^W d x_2^W
f_u (x_1, Q^2 ) f_d (x_2, Q^2 )
\nonumber \\
& \times& P^L_{ W/u }
\left( x_1^W \right) P^L_{ W/d }
\left( x_1^W \right)
\hat{ \sigma } \left( s x_1 x_2 x^W_1 x^W_2 \right)\nonumber \\
&+& (u \leftrightarrow d)
\nonumber \\
\end{eqnarray}
where
in the second line, we have used the fact that
[based on Eq. (\ref{Lemission})] the
average $p^2_T$ of
the longitudinal $V$ is given by $\sim ( 1 - x ) M^2_V \ll
\left( x^W_{ 1, \; 2} E \right)^2$.
Here, $x^W_{ 1, \; 2 } E \sim m^G_n \sim$ TeV is
roughly the energy of the longitudinal $V$
in order to produce an {\em on}-shell KK
graviton\footnote{As mentioned before, the
ratio of signal to background decreases
rapidly as we go outside the resonance region.}.
Hence, we can neglect $p_T$'s in
the parton-level cross-section, i.e., set
$\hat{s} \approx s x_1 x_2 x^W_1 x^W_2$
and integrate over $p_T$'s to obtain
total probabilities, $P^L_{ W/d } (x) = P^L_{ W/ u } (x) \approx
g^2 / \left( 16 \pi^2 \right) \times (1 - x) / x$.
Also,
$f_{ u, d }$ are the $u$, $d$ PDF's;
the $u$ quark (or $W^+$) can come from the first proton and
$d$ quark (or $W^-$) from the second proton or vice versa.
Expressions for contributions from $W/Z_L$ emission from various
other combinations
of quarks and anti-quarks inside the protons
can be similarly obtained.

\end{document}